\documentclass[review]{elsarticle}
\usepackage{lineno,hyperref}
\modulolinenumbers[5]
\journal{Journal of Magnetism and Magnetic Materials}
\usepackage{color}
\usepackage{amsmath}
\begin{document}
\begin{frontmatter}

\title{Electron correlation effects in the exchange coupling at the Fe/CoO/Ag(001) ferro-/antiferro-magnetic interface.}

\author[a]{M. Sbroscia}
\author[a]{A. Verna}
\author[a,2]{G. Stefani}
\author[3,4]{S.R. Vaidya\fnref{q}}
\author[5]{R. Moroni}
\author[5]{F. Bisio}
\author[b,a]{S. Iacobucci}
\author[a]{F. Offi}
\author[6,7]{S. Simonucci}
\author[a]{A. Ruocco}
\author[3]{and R. Gotter
\corref{*}}

\address[a]{Dipartimento di Scienze, Universit\`a degli studi Roma Tre, Via della Vasca Navale 84, I-00146, Rome, Italy}
\address[2]{CNR-ISM sede secondaria Area della Ricerca di Montelibretti, 00015 Monterotondo Scalo}
\address[3]{CNR-IOM, Istituto Officina dei Materiali Area Science Park, I-34149 Basovizza-Trieste, Italy}
\address[4]{Universit\`a di Trieste, Scuola di Dottorato in Nanotecnologie,Trieste, Italy}
\address[5]{CNR-SPIN, Corso Perrone 24, 16152 Genova, Italy}
\address[b]{CNR - Istituto di Struttura della Materia (ISM), Via Fosso del Cavaliere 100, 00133 Rome, Italy}
\address[6]{Universit\`a di Camerino, Camerino, Italy}
\address[7]{INFN, Sezione di Perugia, 06123 Perugia (PG), Italy}
\cortext[*]{gotter@iom.cnr.it}
\fntext[q]{Present address: Department of Physics, Southern Connecticut State University,
501 Crescent St, New Haven, CT 06515, USA}

\begin{abstract}
Angle resolved-Auger-photoelectron coincidence spectroscopy (AR-APECS) has been exploited to investigate the role that electron correlation plays in the exchange-coupling at the ferromagnetic/antiferromagnetic interface of a Fe/CoO bilayer grown on Ag(001). The effective correlation energy U$_\textrm{eff}$, usually employed to assess the energy distribution of core-valence-valence Auger spectra, has been experimentally determined for each possible combination of the orbital (e$_g$ or t$_{2g}$) and the spin (majority or minority) of the two valence electrons involved in the Auger decay. Coulomb and exchange interactions have been identified and compared with the result obtained on the Fe/Ag system. The presented analysis reveals in the Fe/CoO interface an enhancement of the Coulomb interaction for the e$_g$ orbital and of the exchange interaction for the t$_{2g}$ orbital with respect to the Fe/Ag case, that can be associated to the stronger electron confinement and to the exchange coupling between the two layers, respectively.
\end{abstract}
\end{frontmatter}


\section{Introduction}
Developing new kind of magnetic composite devices requires crucial advancements in the comprehension, at the atomic level, of the physics of processes determining interface phenomena, such as those occurring at the ferromagnetic (FM) / antiferromagnetic (AFM) interface. Interface coupling between different magnetic phases plays a role in the behavior of most of the current magnetic and spintronic devices such as tunnel magnetoresistance read heads~\cite{nakatani2018}, magnetoresistive sensor recording media~\cite{piramanayagam2007} or magnetoresistive random access memories (MRAMs)~\cite{chappert2007}, to name a few.

By coupling a FM with an AFM the interaction between the two magnetic phases gives rise to interesting effects, among which the so-called exchange bias (see Nogues \textit{et al.} for a review~\cite{nogues2005}), that amounts to a modification of the magnetization curve of the FM operated by the pinning of the AFM spins. The role that electron correlation plays in this effect is still debated~\cite{pou2002, panda2016, kotliar2006, kanamori1963}.  In general, a detailed understanding of the interplay among band structure, magnetism, and many-body correlations is still in progress~\cite{tushe2018} and needs to be established on firm grounds.
Among the possible FM/AFM bilayers relevant to applications, 3d-FM transition metals (TMs) coupled to their AFM oxides (TMOs) play a pivotal role as they allow to achieve a very good control on the growth process and consequently on their electronic and magnetic structure.
For these reasons, Fe/TMO bilayers are archetype for investigating FM/AFM interfaces and have been extensively studied over the years~\cite{finazzi2009}. Recently, nuclear resonant scattering~\cite{luches2011, luches2012}, magneto optical Kerr effect~\cite{luches2010, colonna2009, kim2010} or X-ray absorption measurements~\cite{colonna2009,  kim2010, abrudan2008}, have reported complex interfacial properties. The results obtained in refs.~\cite{luches2011, luches2012, giannotti2016} point to the presence of an oxide phase of the FM layer in the interface region, but the role it plays in the magnetic interaction has not been established yet. In such systems, it is particularly important to study the local magnetic configuration of the few atomic layers forming the interface at the early stage of formation of the bilayer. Hence, an ultra thin Fe layer grown on top of a metal oxide is a good test bed to investigate effects induced by the substrate on the ferromagnetic overlayer.

Exploiting the information potential of such a system is experimentally
challenging because it requires tools featuring atomic scale sensitivity to electron correlation combined with elemental selectivity. Not many conventional spectroscopies satisfy both conditions; among them Auger spectroscopy is particularly sensitive to electron correlation but its application to FM/AFM interfaces has a drawback: spectra originated by the metals in FM and AFM phases show broad and almost featureless lineshapes that often overlap in energy.
Auger photoelectron coincidence spectroscopy (APECS) ~\cite{bartynski1996} has proven to overcome this difficulty and to be suitable to investigate in a unique way the Auger lineshape of complex and highly correlated systems~\cite{gotter2007, jiang2001}.  These experiments detect, correlated in time, the photo- and the Auger-electron ejected within the same photoionization event, thus yielding  spectra that are made local by the core photoemission and sensitive to electron correlation in the valence band by the Auger decay whose final state features at least two interacting holes. Furthermore, when the emission angle of the two electrons is taken into account, APECS becomes sensitive to the total spin of the two-hole final state ~\cite{gotter2005,gotter2009}.
With the discovery of the dichroic effects in angle resolved APECS (DEAR-APECS), it has been possible to isolate, in the Auger spectrum of FM and AFM thin films, the individual contributions of specific hole-hole pairing originating from spin polarized bands, allowing a deeper insight into the local magnetic structure of these systems~\cite{dapieve2008,cini2011,gotter2011, gotter2012, gotter2013, gotter2020}.

It is the aim of this paper to apply AR-APECS to unravel the role of electron correlation in the exchange-coupling at FM/AFM bilayers. In order to establish possible connections between correlation and exchange-coupling, the results obtained for the interface where such an effect is observed, i.e. Fe/CoO~\cite{abrudan2008}, are compared with results obtained for a system where it is not present, i.e. Fe/Ag, which is assumed as a reference~\cite{gotter2020}.
The paper is organized as follows: section 2 introduces the experimental methods, the experimental results are presented and discussed in section 3 and section 4 is devoted to the conclusions.

\section{Experimental}
The experiments were carried out with the AR-APECS apparatus at the ALOISA beamline of the ELETTRA synchrotron radiation facility (Basovizza - Trieste, Italy)~\cite{gotter2001}. An iron thin film was deposited on a CoO thicker film grown on an Ag(001) crystal surface. The non-magnetic Ag substrate was chosen because of its reduced lattice mismatch with respect to both Fe and CoO, thus allowing a good pseudomorphic growth of these overlayers. Cycles of 1 KeV Ar$^+$-ion sputtering have been performed to clean the Ag(001) single crystal surface, that resulted free of contaminants at an X-ray photoemission analysis. The Ag surface was then annealed at 750 K to achieve good crystallographic order. The CoO conventional cell (face-centered cubic - fcc - rock salt structure, a$_{\mbox{\footnotesize{CoO}}}$ = 4.26 \AA) grows almost in register with the Ag crystal structure (fcc, a$_{\mbox{\footnotesize{Ag}}}$ = 4.09 \AA) and an initial compressive strain gives way to a relaxed film after few monolayers (MLs)~\cite{csiszar2005, torelli2007}. The CoO film was obtained by reactive deposition of Co atoms, evaporated on the Ag (001) substrate by electron bombardment of a high purity metal rod in a controlled atmosphere with an oxygen partial pressure of 1$\cdot$10$^{-4}$ Pa. In order to avoid the formation of clusters with different crystallographic orientations~\cite{abrudan2008}, the Ag substrate was kept at 470 K during deposition and a 30 minutes post-annealing at 750 K was performed in the same oxygen atmosphere. Crystallinity of the overlayer was checked by means of reflection high energy electron diffraction (RHEED), that displayed the same (001) periodicity of the substrate. Photoemission spectra of the CoO film resulted free from contaminants. The same photoemission measurements were used to evaluate the CoO film thickness by comparing peak intensity of Ag and Co core states, thus calibrating the deposition rate~\cite{luth, fadley}. A 25 ML-thick CoO film was employed for the subsequent growth of the Fe epilayer.
A 2 ML-thick Fe film was epitaxially grown on top of the CoO overlayer, by electron bombardment of a high purity metal rod. The surface quality was monitored by RHEED, photoemission spectroscopy and the X-ray absorption at the Fe L$_{23}$ edge. The Fe thickness was estimated by using  evaporation conditions (time and rate) identical to those used  to achieve two complete RHEED oscillations~\cite{luth} in a previous experiment performed on  Fe/Ag(001) with the same experimental setup ~\cite{gotter2020}.

\begin{figure}[h!]
      \begin{centering} \includegraphics[width=1\textwidth]{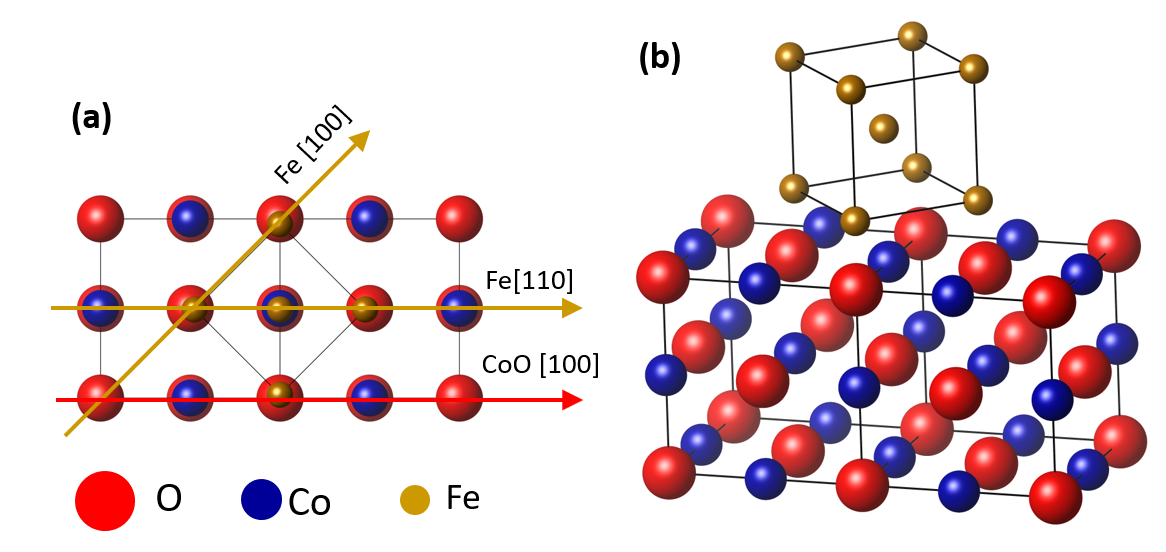}
      \caption{Top (a) and oblique (b) view of the Fe/CoO (001) interface showing the Fe cubic cell rotated by 45$^{\circ}$ with respect to the CoO cell, according to the model presented in the Fig. 2 of ref.~\cite{brambilla2008}. The four atoms defining the lower face of the Fe cell sit on top of the underlying O ions. See text for further details.}
      \label{Fig1}
\end{centering}
\end{figure}

The Fe film grows epitaxially on the CoO substrates with the conventional crystalline cell (body-centered cubic - bcc - a$_{\mbox{\footnotesize{Fe}}}$ = 2.87 \AA) rotated by 45$^{\circ}$ with respect to the cells of underlying CoO and Ag~\cite{abrudan2008,  brambilla2008, li1990}. In Fig.~\ref{Fig1} a model of the Fe/CoO  interface shows that the diagonal of the rotated Fe unit cell matches the edges of the CoO unit cell, in accordance with the simple model presented by Brambilla \textit{et al.}~\cite{brambilla2008}. In the figure the four atoms defining the lower face of the Fe cell sit on top of the underlying O ions. Even though a bond between Fe atoms and O ions of the CoO film is expected, real interfaces between Fe and simple oxides can be  more complex and the formation of iron oxides has been demonstrated both by experiment and calculations~\cite{colonna2009, giannotti2016, valeri2007}.

A Curie temperature T$_C$ well above RT is found in literature for a 2 ML thick Fe films grown on Ag (001)~\cite{lang2006, qiu1993, stampanoni1987}. The N\'eel temperature (T$_N$) of the  25 ML-thick CoO film equals the bulk one, i.e. T$_N$ = 290 K~\cite{abrudan2008}. During all measurements, the sample temperature has been kept constant at 170 K, well below the critical temperatures of both the FM and AFM phases,  assuming for the Fe layer grown on CoO a $T_C$ close to that of the Fe/Ag system with the same Fe thickness.
It is well established that the orientation of the magnetization in thin films strongly depends on thickness, temperature and thermal treatment. Annealed samples usually display a reorientation of the magnetization from out-of-plane to in-plane increasing thickness and/or temperature~\cite{qiu1993}. For non-annealed samples, no out of plane anisotropy is found above 100 K for thicknesses ranging from 0.8 to 10 ML~\cite{stampanoni1987}. However, this experiment relies only on the quantisation axis identified by the electric field of the impinging linearly polarized light. Therefore, being not necessary to know the sample magnetization $M$ to interpret the experimental result, no external magnetic field was applied to the investigated samples before or during the measurements.

\begin{figure}[h!]
\begin{centering}
\includegraphics[width=0.8\textwidth]{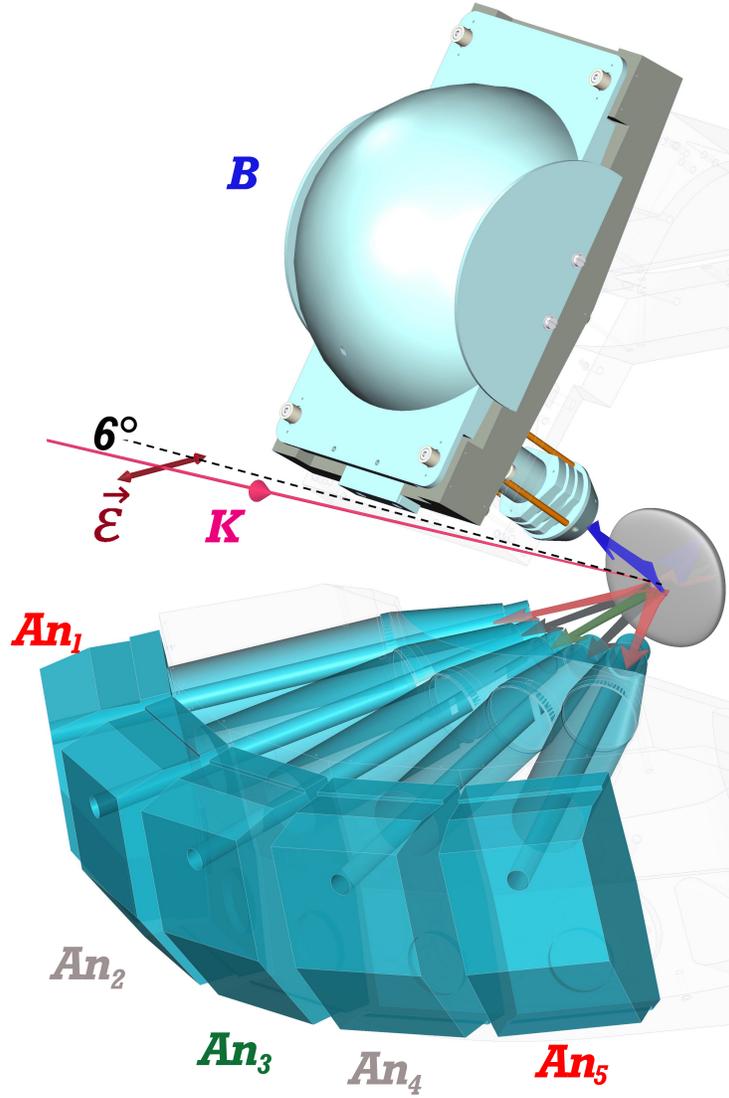}
      \caption{Experimental setup: monochromatic linearly polarized photons with h$\nu$=250 eV impinge onto the sample with a grazing incidence angle of 6$^{\circ}$. The sample normal lies in the polarization - momentum $\boldsymbol{\epsilon K}$ plane, hence a \textit{p}-polarization scheme is set. Fe 3p photoelectrons are collected in the $\boldsymbol{\epsilon K}$ plane at different polar angles by the energy-single-channel analyzers An1-An5, while the Auger electrons are collected by the multichannel analyzer B, 38$^{\circ}$ apart from the $\boldsymbol{\epsilon K}$ plane. Considering that $\boldsymbol{\epsilon}$ acts as a quantization axis, with this arrangement different kinematics can be accessed, sensitive to final state with antiparallel spins (AS) or  parallel spins (PS) of the two emitted electrons. See text for further details.}
\label{Fig2}
\end{centering}
\end{figure}

\begin{figure}[h!]
\begin{centering}
\includegraphics[width=1\textwidth]{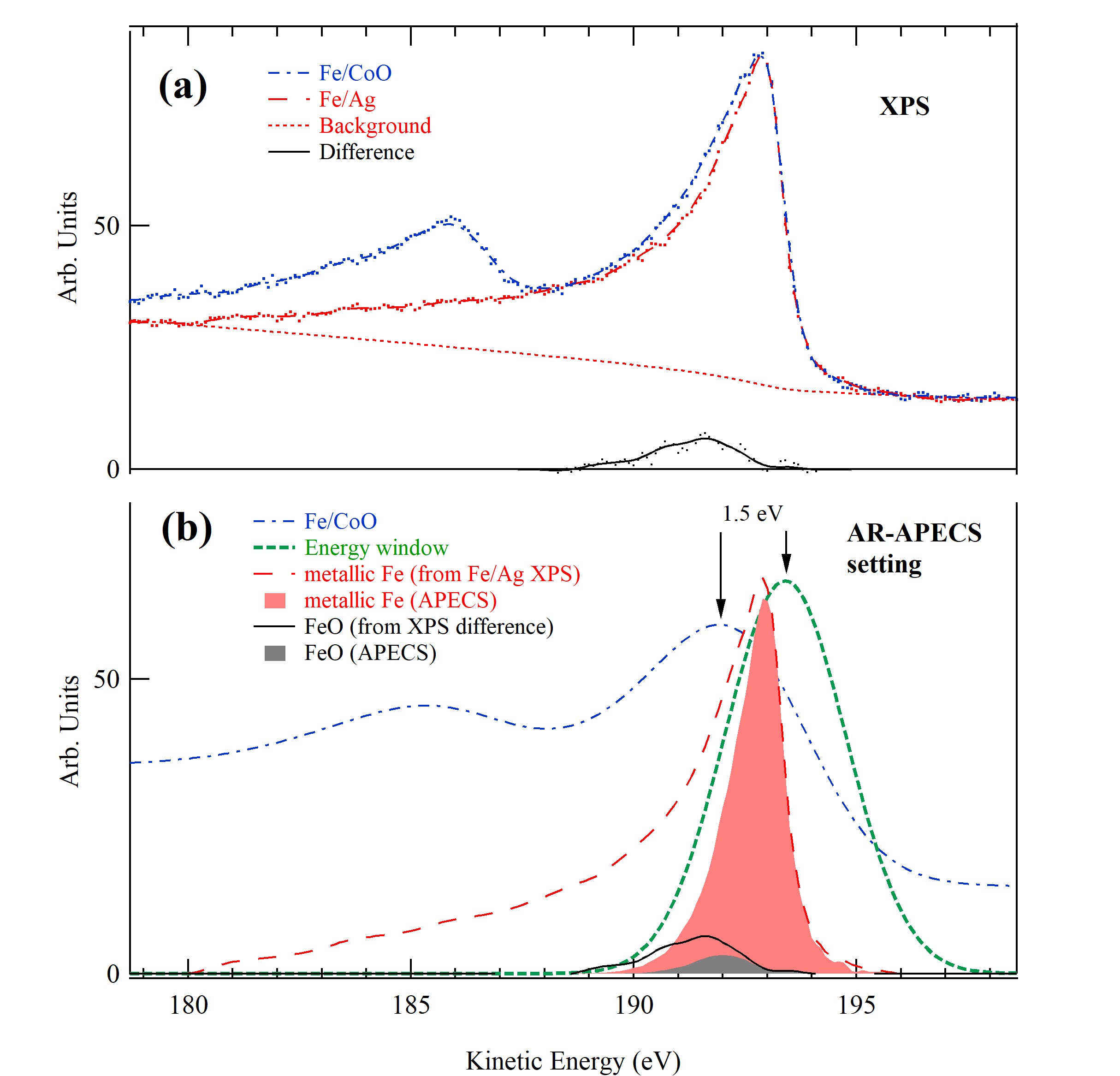}
\caption{(a): difference (continuous black line) between Fe 3p high resolution (0.2~eV) XPS spectra from the two samples Fe/Ag (red dashed-dotted line) and Fe/CoO (blue dashed line), after alignment and normalization at their maximum intensity; this difference spectrum is calculated only in the 189--200~eV energy region, ignoring the portion of the spectra at lower kinetic energy where the Co 3p peak is present for the Fe/CoO interface; the red dotted line is the integral background for Fe/Ag case. (b): in the AR-APECS measurement, the energy window of the photoelectron analyzers (green dashed line) with 3.2 eV energy resolution, has been set 1.5 eV at higher kinetic energy with respect to the Fe 3p maximum intensity measured with the same resolution (blue dashed-dotted line); in such a way, the 3p photoelectrons accepted in the coincidence detection due to the metallic iron (red filled peak) and due to the oxidized iron (grey filled peak) have been estimated (see text for details). The contribution from FeO results to be equal to 10\% in XPS spectra and 5\% in APECS spectra.}
\label{Fig3}
\end{centering}
\end{figure}

The AR-APECS apparatus employed in this work, is discussed in detail elsewhere~\cite{gotter2001}, except for the energy-multichannel electron analyzer that replaced two of the seven energy-single-channel analyzers, previously used.
Schematically, the experimental apparatus consists of six electron energy-analyzers mounted on two different frames: five single-channel, with optical axes lying in the plane defined by the momentum $\boldsymbol{K}$ and the linear polarization $\boldsymbol{\epsilon}$ of the photon beam, and one multichannel positioned 38$^{\circ}$ away from such a plane (see Fig.~\ref{Fig2}).
A photon beam set to h$\nu$ = 250 eV impinged onto the sample surface in \textit{p}-polarization and at grazing incidence, the surface normal being 6$^{\circ}$ away from $\boldsymbol{\epsilon}$. In the experiment, Fe 3p photoelectrons were collected in the $\boldsymbol{\epsilon K}$ plane by the five single-channel analyzers (An$_i$ in Fig.~\ref{Fig2}); each of them encompassed a different polar angle with respect to $\boldsymbol\epsilon$ i.e. 0$^{\circ}$, $\pm$18$^{\circ}$ and $\pm$36$^{\circ}$, thus defining different AR-APECS kinematics.
The energy resolution of these five analyzers was set to 3.2 eV and the accepted energy window was detuned 1.5 eV towards kinetic energies higher than the 3p photoelectron peak maximum, in order to collect mainly the three photoemission lines closely packed at the high kinetic energy side of the 3p photoemission sextet~\cite{Rossi1994, VanDerLaan1995}. In doing this, the signal from oxidized iron, which contributes with a component chemically shifted 1.9 eV at lower (higher) kinetic (binding) energy ~\cite{McIntyre1977, Sinkovic1990}, is lowered as explained as follows.
The quantitative effect of such a setting is explained in Fig.~\ref{Fig3}. In Fig.~\ref{Fig3}a, the Fe 3p high resolution (0.2 eV) X-ray photoelectron spectroscopy (XPS) spectrum of the Fe/CoO interface is compared with the one measured for the Fe/Ag interface~\cite{gotter2020}, after alignment and normalization to their maximum intensity; their difference, shown by the black continuous line and which is interpreted as an evidence for the formation of iron-oxygen bonds at the Fe/CoO interface, provides an estimate of the contribution to the Fe~3p peak  intensity due to iron atoms bonded to oxygen, which results less than 10\% of the  contribution ascribable to metal atoms.
In Fig.~\ref{Fig3}b, the Fe 3p peak measured with 3.2 eV energy resolution, as used in the APECS setting, is shown together with the energy window accepted by the photoelectron analyzers (3.2 eV FWHM), which is set 1.5 eV above the Fe 3p maximum intensity. Multiplying the energy window by the two Fe 3p contributions (the metallic one as derived by the Fe/Ag XPS after background subtraction, and the oxide one derived by the difference spectrum) allows to quantify the contribution to the AR-APECS spectra due to oxidized iron to be less than 5 \% or, in other words, the contribution from metallic iron is dominant in AR-APECS spectra.

Fe M$_{3}$VV (that is a M$_{3}$M$_{4,5}$M$_{4,5}$ super Coster-Kronig decay) Auger electrons, were collected at an angle of 38$^{\circ}$ off the $\boldsymbol{\epsilon K}$ plane by the multichannel analyzer (B in Fig.~\ref{Fig2}), set to provide an energy resolution of 1 eV while scanning the full spectrum. Taking into account the 3p$_{3/2}$ core-hole lifetime broadening ($\Gamma$ = 0.48 eV~\cite{nyholm1981}), and the many-body effects accompanying the Auger transition~\cite{verdozzi2001}, a total intrinsic broadening of 1.6 eV FWHM is estimated for the Auger measurements.

Depending on the different collection angles selected by the analyzers, a moderate selection of partial waves with a specific $m$ quantum number associated to the two emitted electrons, is achieved. These are spherical harmonics defined with respect to $\boldsymbol{\epsilon}$ whose amplitude is modulated by diffraction from the crystal lattice~\cite{gotter2009,dapieve2008}. Namely, the wavefunction of electrons emitted within a cone centered around $\boldsymbol{\epsilon}$ and with an aperture angle of about 20$^{\circ}$ has almost pure $m = 0$ character, while electrons emitted outside of this cone, have predominantly character $m = 1$ for photoelectrons and $m \geq 1$ for Auger electrons (the higher the deviation angle from $\boldsymbol{\epsilon}$ the higher the dominant $m$).
The angular selection operated on the photoelectron implies that only the subset of core-hole states with the selected magnetic quantum number $m$ are involved in the following step. The detection in time coincidence of the subsequent Auger electron will select autoionizing events that originate not from a statistical population of core-hole states, but rather from that specific subset of them, i.e. from an aligned core state. By selecting in angle the Auger electron, a further constraint on the allowed $m$ for the Auger wave function is imposed. Taking into account the two discriminated $m$ values, and the selection rules for photoemission and Auger processes, a selectivity on specific final states is established, which has been extensively discussed in previous works~\cite{gotter2003, gotter2005, dapieve2007}.  
In short, if the two electrons are ejected close to the polarization $\boldsymbol{\epsilon}$, ($m=0$ for both) the combination of the selection rules with the Pauli exclusion principle dictates that the two electrons are ejected with opposite spin. Therefore, only final states associated to the emission of an electron pair with antiparallel spin can be ascribed to the observed AR-APECS spectrum. On the contrary, when electron detected far from the $\boldsymbol{\epsilon}$ direction are involved, larger values of $m$ are favored, thus allowing also the emission of two electrons with parallel spin.
In general, the relative weight of antiparallel spin versus parallel spin emission can be expressed as a function of $|\Delta m| = |m_1-m_2|$, where $m_1$ and $m_2$ are the $m$ values of the two emitted electrons; the higher $|\Delta m|$, the higher the probability of parallel spin emission~\cite{gotter2005}.
In this experiment, for Auger electrons emitted 38$^{\circ}$ from the $\boldsymbol{\epsilon K}$ plane, a mix of $m = 0$, 1 and 2 components of the dominant $f$ Auger electron wavefunction is detected. For the photoelectron collected by the central analyzer (An$_3$ in Fig.~\ref{Fig2}) along the polarization vector direction, mostly described by a $d$ wavefunction, the $m = 0$ component dominate and relatively low values of $|\Delta m|$ are involved.
In analogy to previous AR-APECS experiments ~\cite{cini2011,gotter2020}, the analyzers pairs selecting photoelectrons emitted close to the polarization vector will be termed as antiparallel spin configuration (AS) , while pairs with the photoelectron far from $\boldsymbol{\epsilon}$ will be termed parallel spin configuration (PS).
The coincidence count rate for these experiments was of the order of 8.6 $\cdot$ 10$^{-2}$ counts per second, thus nearly 65 h of integration time have been required to achieve a good statistic. Due to the reactivity of the iron surface, sample quality has been constantly monitored and to prevent oxidation a new sample has been prepared every 12 h of beam exposure.

\section{Results and Discussion}
A thorough analysis of the conventional Auger electron spectroscopy (AES) is reported as supplementary material ~\cite{supplement}, together with the implementation of a Fe-O molecular model~\cite{taioli2009,taioli2010,taioli2009b}, used to quantify the effect on the Fe M$_3$VV lineshape of the iron atoms bonded to the oxygen. The main result are:
(i) trying to infer the  M$_3$VV lineshape from conventional AES is an unreliable procedure that leads to the wrong conclusion that electron correlation is irrelevant and then the lineshape can be fitted by the self-convolution of the valence band density of states;
(ii) the effect due to a small amount of iron oxide at the buried interface (less than 5\% as probed by XPS) is minimal and negligible in the APECS spectra, which are in turn more surface sensitive with respect to non coincident XPS and AES spectra ~\cite{werner2005}.

\begin{figure}[h!]
\begin{centering}
\includegraphics[width=1\textwidth]{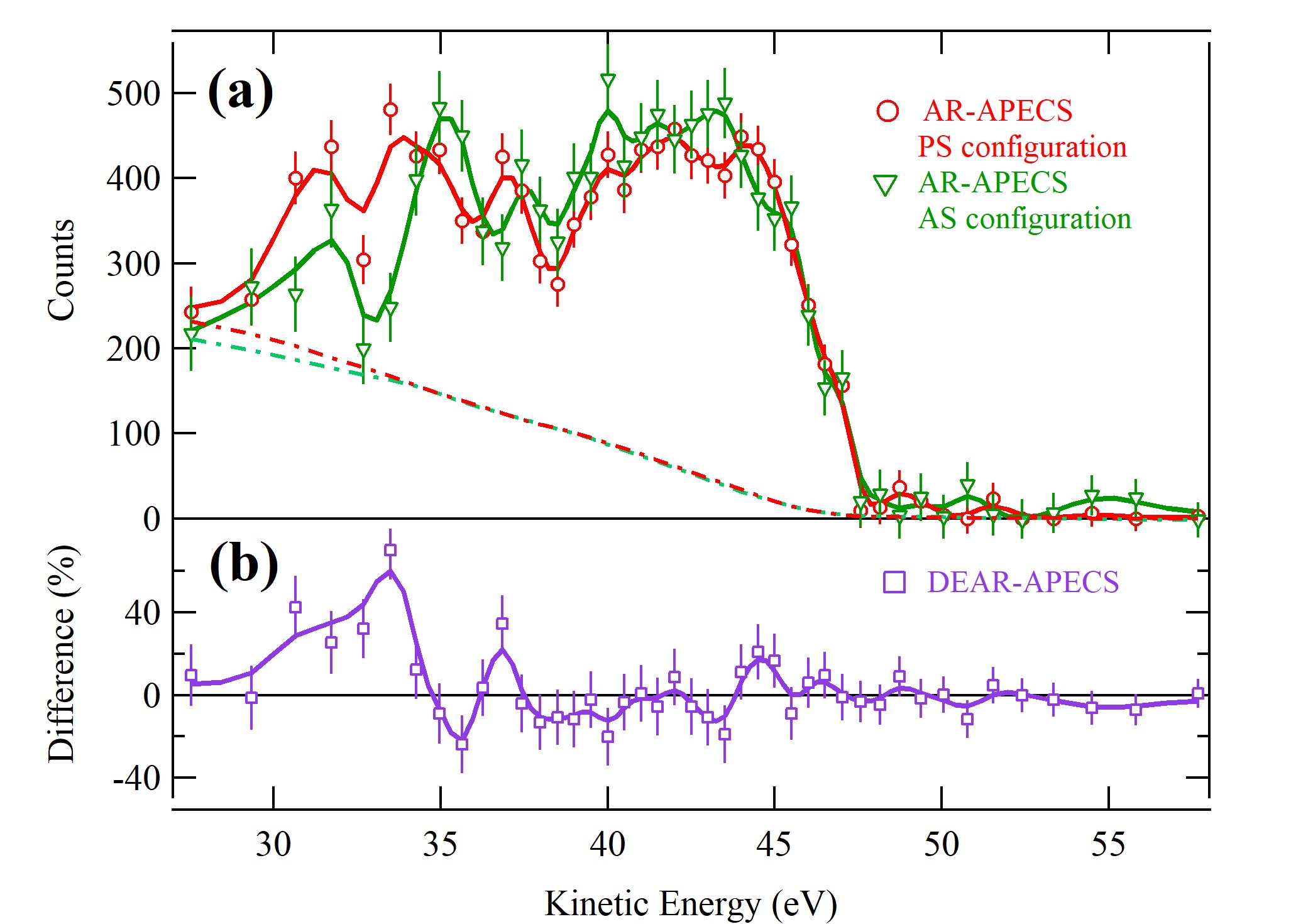}
\caption{(a) AR-APECS spectra from Fe/CoO measured in PS (red circles and line as guide for the eyes) and  AS (green triangles and line) configurations, are shown together with their estimated background due to intrinsic secondary electrons (dashed-dotted lines). (b) the dichroic effect in AR-APECS (DEAR-APECS) (violet open squares) is defined as the difference between PS and AS spectra divided by the their average over the energy interval from 30 to 50 eV: it is therefore a difference curve expressed in percent of the average integrated intensity}; the continuous line is a guide for the eyes.
\label{Fig5}
\end{centering}
\end{figure}

The angle-resolved AR-APECS measurements are reported for the Fe/CoO case in Fig.~\ref{Fig5}a, as measured in the AS and PS configuration, respectively. The dichroic effect associated to AR-APECS (DEAR-APECS) reported in Fig.~\ref{Fig5}b, that is the difference between the Auger lineshapes collected in PS and AS configurations, is more relevant in the lower energy section of the spectrum. Previous AR-APECS investigations on Fe thin films ~\cite{gotter2012, gotter2020} have associated features in the low-energy region of the spectrum with resonant hole-hole final states.
In particular, the Fe Auger spectrum has been accurately investigated in the case of a 2 ML Fe film directly deposited on Ag(001) and its shape was found to be composed by a manifold of hole-hole resonances well accounted by the Cini-Sawatzky (CS) theory ~\cite{cini1976, cini1977, sawatzky1977} with an effective electron correlation energy U$_\textrm{eff}$ that resulted to be dependent upon the pairing, in the two-hole final state, of spin and of ligand field orbitals~\cite{gotter2020}. The following data analysis will proceed in a similar way in order to compare final results. In order to take into account majority and minority spin sub-bands, as well as e$_g$ and t$_{2g}$ ligand field orbitals, the same density of states (DOS) calculated by Rhee~\cite{Rhee2005} for Fe/Ag is used here. The very similar APECS and AR-APECS spectra of the two bilayers Fe/Ag and Fe/CoO, together with the fact that the Auger signal from iron atoms bonded to oxygen amounts to less than 5\%, lead to the conclusion that the same DOS can be used as input of the CS model for simulating also Fe/CoO Auger spectra.
Assuming four sub-bands, associated to e$_g$ and t$_{2g}$ orbitals, each with majority and minority spin, there are ten possible combinations for the creation of two holes. Three lead to the emission of two electrons with parallel spin $\uparrow\uparrow$, four with antiparallel spin $\uparrow\downarrow$, and three with parallel spin $\downarrow\downarrow$. These latter will be neglected due to the low $d$ electron population of the e$_g$ and t$_{2g}$ orbitals~\cite{Rhee2005}, indeed inhibiting the $\downarrow\downarrow$ decays where both holes are created in the same ligand field orbital and allowing only the e$_g^\downarrow$t$_{2g}^\downarrow$ decay.

\begin{figure}[h!]
\begin{centering}
\includegraphics[width=01\textwidth]{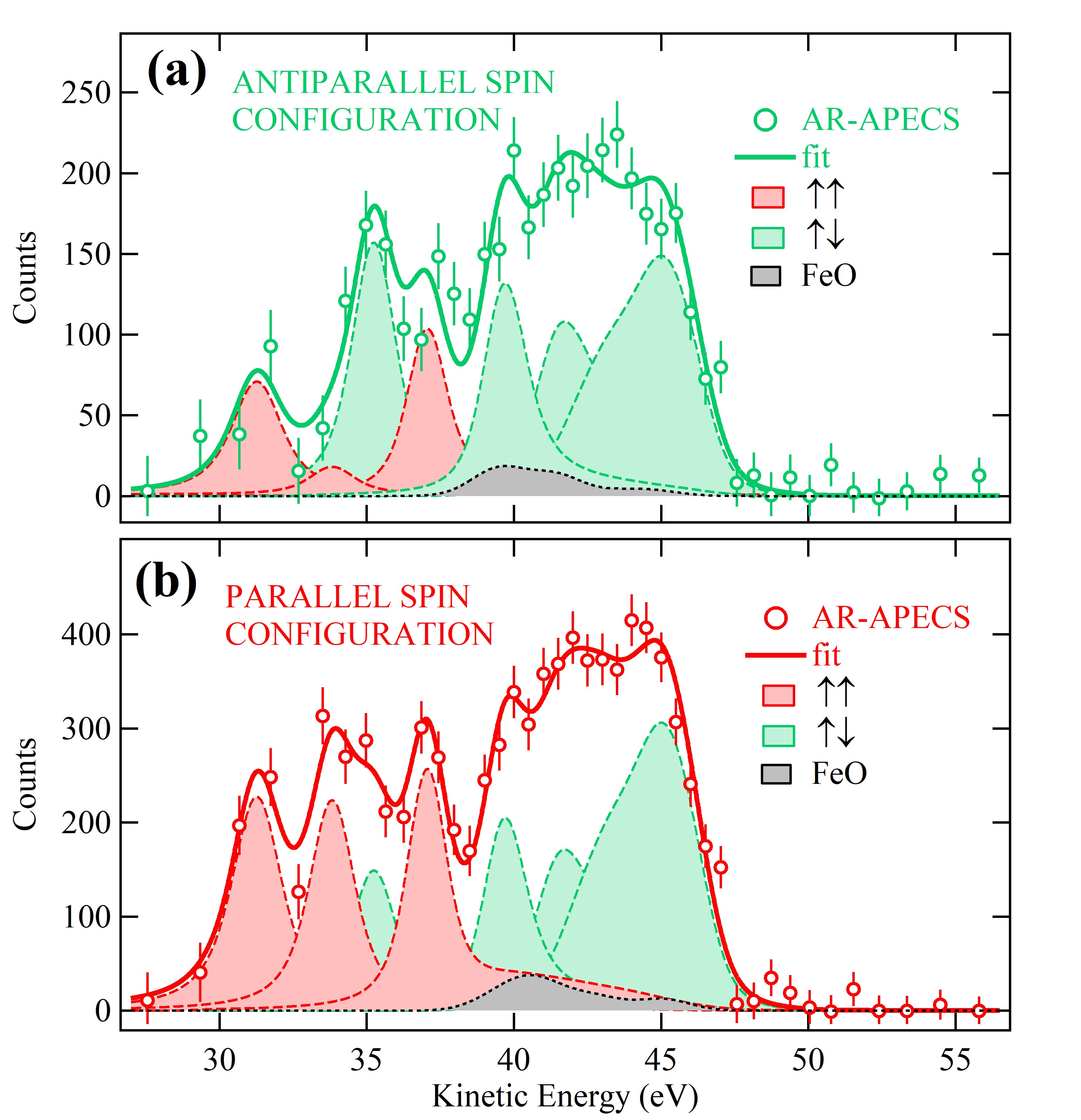}
\caption{Fit of the experimental AR-APECS spectra from Fe/CoO measured in AS (a) and PS (b) configurations after background subtraction, green and red continuous lines, respectively.  The hole-hole components, three with parallel spin character (red filled peaks), and four with antiparallel spin character (green filled peaks), as well as the Fe-O molecular model (grey filled peak) to account for the iron atoms bonded to the oxygen, are shown for both AS and PS spectra.}
\label{Fig6}
\end{centering}
\end{figure}

The background-subtracted AR-APECS spectra are shown in Fig.~\ref{Fig6}. The dichroic effect enables to assign unambiguously the spin character of the individual spectral components; similarly to what found on Fe/Ag~\cite{gotter2020}. There are three parallel spin peaks (red filled curves in Figure~\ref{Fig6}) which are more intense in the PS experimental configuration, and four antiparallel spin peaks (green filled curves in Figure~\ref{Fig6}) which are dominant in the AS configuration.
The two AR-APECS spectra have been fitted simultaneously by a function in which the Cini formula is applied to each possible pairing of the individual spin and orbital components of the theoretical DOS, each one with its specific U$_\textrm{eff}$ value, plus the contribution of the Fe-O model. In particular, in order to properly include the molecular model we have taken into account the dependence of the photoionization cross section for the 3$\sigma$ and 3$\pi$ core-hole states on the angle between $\boldsymbol{\epsilon}$ and the photoemission direction, resulting in two slightly different curves (dashed curves in Figures~\ref{Fig6}a and~\ref{Fig6}b).
The result of the least square fitting procedure, corresponds to the continuous green and red lines in Fig.~\ref{Fig6}a and~\ref{Fig6}b, respectively. In Table~\ref{tab1} the position in kinetic energy and the associated U$_\textrm{eff}$ energies for each hole-hole resonance, as determined by the fitting procedure, are reported, labelled with the two orbital and the two spin involved in the decay, and sorted by decreasing value. The values found for the Fe/Ag bilayer~\cite{gotter2020} are also reported in parenthesis, for comparison.
The molecular calculation contributes to the model lineshape only in a small region of the weakly correlated part of the spectrum. This means that the formation of a partially oxidized Fe layer is properly taken into account, and it does not play any role in the low kinetic energy section of the spectrum where the sharp resonant features are singled out by AR-APECS.
The similar behavior of Fe/CoO and Fe/Ag AR-APECS spectra suggests that the resonant features originate from the same combinations of spin and orbitals in the two hole final states, but with electron-correlation interactions of different strength. It has been shown~\cite{treglia1981} that, even in case of partially filled bands, the energy separation between  the top of the leading edge of the uncorrelated part of the spectrum and the sharp hole-hole resonance is linearly dependent upon the electron correlation energy U$_\textrm{eff}$ characteristic of the specific hole pair involved. Hence, the difference in energy between analogous transitions as measured in Fe/CoO and Fe/Ag~\cite{gotter2020} corresponds to the variation of U$_\textrm{eff}$ experienced by Fe valence holes in going from the non magnetic (Ag) substrate to the antiferromagnetic (CoO) one. 

\begin{table*}[h!]
\centering
\begin{center}
\begin{tabular}{lll|rl}
\hline \hline
Hole-hole &  E$_\textrm{kinetic}$          &  U$_\textrm{eff}$           & Significative energies  & Significance and   \\
pairing   &  [eV] &  [eV$\pm$0.1 eV] & [eV$\pm$0.2 eV] & $E^\alpha(\beta)$ (see text) \\
\hline \hline                 
$e_g^\uparrow e_g^\uparrow$ & 31.3 & 11.9 &
U$_{e_{g}^\uparrow e_{g}^\uparrow}$-U$_{e_{g}^\uparrow e_{g}^\downarrow}$ = 3.4 & $e_{g}$ spin flip $E^{e_{g}}(e_{g})$\\
 &  & (11.0) & (3.5)\\

$t_{2g}^\uparrow t_{2g}^\uparrow$ & 33.9 & 9.2 &
U$_{t_{2g}^\uparrow e_{g}^\uparrow}$-U$_{t_{2g}^\uparrow e_{g}^\downarrow}$ = 4.2 & $e_{g}$ spin flip $E^{e_{g}}(t_{2g})$\\
 &  & (8.7) & (4.2)\\

$e_g^\uparrow e_g^\downarrow$ & 35.0 & 8.5 &
U$_{t_{2g}^\uparrow t_{2g}^\uparrow}$-U$_{t_{2g}^\uparrow t_{2g}^\downarrow}$ =$\boldsymbol{5.1}$ & $t_{2g}$ spin flip $E^{t_{2g}}(t_{2g})$\\
 &  & (7.5) & ($\boldsymbol{4.4}$)\\

$e_g^\uparrow t_{2g}^\uparrow$ & 37.2 & 5.8 &
U$_{e_{g}^\uparrow t_{2g}^\uparrow}$-U$_{e_{g}^\uparrow t_{2g}^\downarrow}$ = $\boldsymbol{5.8}$ & $t_{2g}$ spin flip $E^{t_{2g}}(e_{g})$\\
 &  & (5.2) & ($\boldsymbol{5.2}$)\\

$t_{2g}^\uparrow t_{2g}^\downarrow$ & 39.9 & 4.1 &
U$_{e_{g}^\uparrow e_{g}^\downarrow}$ = $\boldsymbol{8.5}$ & $e_{g}e_{g}$ Coulomb \\
 &  & (4.3) & ($\boldsymbol{7.5}$)\\

$t_{2g}^\uparrow  e_g^\downarrow$ & 42.4 & 1.6$\pm$0.2 &
U$_{t_{2g}^\uparrow t_{2g}^\downarrow}$ = 4.1 & $t_{2g}t_{2g}$ Coulomb\\
 &  & (1.0$\pm$0.2) & (4.3)\\

$e_g^\uparrow t_{2g}^\downarrow$& 44.6 & 0.0 & \\
 &  & (0.0) & \\

\hline \hline
\end{tabular}
\end{center}
\caption{Electron-correlation energies U$_\textrm{eff}$ , associated to the Auger features singled out by AR-APECS measurements, are listed in the third column for Fe/CoO and Fe/Ag (in parenthesis~\cite{gotter2020}), for each possible hole-hole combination (first column), but neglecting the  three $\downarrow\downarrow$ contributions (see text); approximate values of the kinetic energy position of each component are also reported in the second column, to allow for their recognition in Fig.~\ref{Fig6}. In the fourth column spin-flip energies are expressed in terms of energy difference between final states having the same orbital pairing but associated to parallel and antiparallel spin of the two emitted electrons; Coulomb energies are also listed (see text); in bold, the values presenting a significative difference between Fe/CoO and Fe/Ag are highlighted. In the $E^\alpha(\beta)$ terms, used in the text, the superscript indicates the orbital of the flipped spin  while in parenthesis the spectator hole is indicated. In spite of the relatively large error bars of the individual experimental points, the uncertainties of the correlation energies U turns out to be small because the two AR-APECS spectra, in the AS and PS configurations, are fitted simultaneously}
\label{tab1}
\end{table*}

Similarly to what was found for Fe/Ag~\cite{gotter2020}, the variation in U$_\textrm{eff}$ recorded between final states involving identical orbitals but different spin configurations (parallel or antiparallel) are understood as the spin-flip energy of those states. It is therefore possible to identify the effects of the Coulomb and the exchange interactions, separately.
For instance, the difference between the observed energies of the $e_g^\uparrow e_g^\uparrow$ and $e_g^\uparrow e_g^\downarrow$ final states, equal to 3.4 eV, is the spin-flip energy of an $e_g$ electron, paired with an $e_g$ spectator hole. In this way the correlation energy associated to the final state $e_g^\uparrow e_g^\uparrow$ (U$_\textrm{eff}$=11.9 eV for Fe/CoO) is understood as due to the Coulomb interaction derived by the correlation energy of the $e_g^\uparrow e_g^\downarrow$ final state (U$_\textrm{eff}$=8.5 eV), plus the spin-flip energy.
The Coulomb and exchange contributions to the Auger spectra, singled out for each hole-hole orbital pairing, are listed in the fourth column of Table~\ref{tab1}, where in bold the most significative differences between the two interfaces are highlighted.
Within the experimental uncertainty, it is possible to assess that, in going from Fe/Ag to Fe/CoO, the Coulomb interaction increases by 1 eV for the e$_g$ orbitals while remains the same for the t$_{2g}$ ones; on the contrary, the exchange interaction is enhanced by 0.7 eV for the t$_{2g}$ orbitals due to the FM/AFM coupling, while it remains unchanged for the e$_{g}$ electrons.

In order to correlate this result with the FM/AFM exchange coupling is helpful to recall the concept of spin-flip energy used to define the exchange-splitting in the Stoner model. Single electron spin-flip energies (Stoner excitations) are dominated by intra-atomic interactions, but they are also affected by inter-atomic interactions, which are usually described by Heisenberg-like exchange interactions $J_{ij}\;S_i\;S_j$ in terms of atomic total moments $S$ ~\cite{sthor2006}.
Let's define the spin-flip energies $E^{\alpha}(\beta)$, with $\alpha,\beta = e_g, t_{2g}$ (see table \ref{tab1}), where the superscript indicates the orbital of the flipped spin while in parenthesis the spectator hole (the second hole created in the Auger decay) is indicated. We have for instance for the $t_{2g}$ electrons
$$E^{t_{2g}}(t_{2g}) =\textrm{U}_\textrm{eff}(t_{2g}^\uparrow t_{2g}^\uparrow) -\textrm{U}_\textrm{eff}(t_{2g}^\uparrow t_{2g}^\downarrow)$$ 
and $$E^{t_{2g}}(e_{g}) =\textrm{U}_\textrm{eff}(e_g^\uparrow t_{2g}^\uparrow) -\textrm{U}_\textrm{eff}(e_g^\uparrow t_{2g}^\downarrow)$$
which are the energies necessary to flip the spin of a $t_{2g}$ electron, in two different electronic configurations of the doubly ionized Auger final state: one with a $t_{2g}$ and one with an $e_{g}$ spectator hole, respectively.
The total single electron spin-flip energy $E^{\alpha}(\beta)$, can be written as the sum of an intra-atomic term and an inter-atomic one
$$E^\alpha(\beta)=E_{atom}^\alpha(\beta)+E_{inter}^\alpha(\beta)$$
The inter-atomic term has the Heisenberg-like form
$$E_{inter}^\alpha(\beta)=\sum_j J_{ij}\;\Delta S_i\; S_j = \sum_j J_{ij}\ S_j 
$$
where the sum is over the first neighbor-atoms and describes the variation of the exchange energy due to the spin flip of a single electron at the atomic site $i$, which brings the total atomic spin of from S$_{\textrm{initial}}$ to S$_{\textrm{final}}$, with $\Delta S_i = S_{\textrm{final}} -S_{\textrm{initial}} = 1$.
The contributions are different for atoms in the top or in the bottom iron layer, specifically
\begin{equation}
E_{\textrm{top}}^\alpha(\beta) = E_{\textrm{atom}}^\alpha(\beta) + \sum_{Fe}^{bottom} J_{\textrm{Fe-Fe}} \; S_\textrm{{Fe}}\, 
\label{top}
\end{equation}
(the first neighbor-atoms of a top iron atom are indeed located in the bottom layer), and
\begin{equation}
E_{\textrm{bottom}}^\alpha(\beta) = E_{\textrm{atom}}^\alpha(\beta) + \sum_{Fe}^{top} J_{\textrm{Fe-Fe}}\; S_{\textrm{Fe}} + \sum_{Co}^{substrate} J_{\textrm{Fe-Co}}\; S_{\textrm{Co}}
\label{bottom}
\end{equation} 
where the sum over the first neighbor-atoms of a bottom iron atom includes a sum over the overlying Fe atoms of the top layer and a sum over the underlying Co atoms of the AFM substrate.
By assuming that: (i) $J_{\textrm{Fe-Ag}}=0$; (ii) $S_{\textrm{Co}}=3/2$, as dictated by the ground state $^4F_{9/2}$ of the $3d^7$ cobalt electronic configuration; (iii) the intra-atomic contribution $E_{\textrm{atom}}^\alpha(\beta)$ does not change significantly when considered for the two different substrates; it follows that, when calculating the difference in the spin-flip energy $\Delta E^\alpha(\beta)$ between the Fe/Ag and Fe/CoO systems, the contributions of equation (\ref{top}) cancel out while those of equation (\ref{bottom}) simply reduce to
\begin{equation}
\Delta E^\alpha(\beta) = E^\alpha(\beta)_{\textrm{Fe/CoO}} -  E^\alpha(\beta)_{\textrm{Fe/Ag}} = \sum_{Co}^{bottom} J_{\textrm{Fe-Co}}\;sign(S_{Co}) \frac{3}{2}
\label{coupling}
\end{equation}
The sign of the Co atoms spin has been introduced to include the antiferromagnetic alignment which may give spins of opposite sign in the same coordination sphere.
From the experiment we have that the left side of equation (\ref{coupling}) is zero for  $\alpha = e_g$ and $\approx 0.7$ eV for $\alpha = t_{2g}$, while it is reassuring the fact that it does not depend on the spectator hole $\beta$. This means that the exchange interaction $J_{\textrm{Fe-Co}}$ also depends on the orbital $\alpha$. In other words, the experimentally measured quantities represented in the left side of equation (\ref{coupling}) open a degree of freedom internal to the electronic structure which is not contemplated in the Heisenberg-like formalism dealing with total atomic moments, highlighting the fact that $e_g$ and $t_{2g}$ electrons play different roles in the magnetism of transition metals, in particular when systems with different properties are put together in interaction as in complex electromagnetic nano-devices.
The energy amount in (\ref{coupling}) is finally directly connected to the second term of the Mauri model (formula (1) in~\cite{mauri1987}) describing the exchange interaction $J_{\textrm{FM-AFM}}$ between the FM spins and the AFM spins pinned at the interface, which is supposed to be smaller with respect to the bulk exchange stiffness of both FM and AFM materials. Furthermore, in principle, for perfect interfaces with spin-compensated AFM surfaces the sum in equation (\ref{coupling}) would be zero, but, indeed, a non-zero amount of uncompensated pinned spins for spin compensated surfaces~\cite{malozemoff1988}, as well as a very low number of pinned spins in uncompensated surfaces~\cite{ohldag2003}, have been understood for real interfaces. The sum in equation (\ref{coupling}) would be then substituted with an effective number of pinned spins.

In conclusion, the present experiment provides a local microscopic measure, in the FM magnetic thin film interfaced with the AFM substrate, of the single electron spin-flip energy, which appears to be very sensitive to the FM/AFM exchange coupling. As a matter of fact, at 170 K, the exchange bias field on the CoO(001) surface is very small: one order of magnitude smaller than in the CoO(111) one, due to the (partial) spin compensation on the CoO(001) surface~\cite{mlynczak2013}.
The results of this investigation allow to conclude that the exchange coupling determines an increase of the average exchange interaction in iron as probed by AR-APECS, only on the more itinerant t$_{2g}$ orbitals, while the e$_{g}$ ones are almost unaffected. On the other hand, the enhancement of the Coulomb interaction observed on the more localized e$_{g}$ orbitals~\cite{katanin2010} in the Fe/CoO interface  with respect to the Fe/Ag one, can  be attributed to an enhanced localization of the e$_{g}$ electrons experiencing the proximity with the insulating substrate, with respect to the metal contact with Ag.

\section{Conclusions}
AR-APECS spectroscopy has allowed to achieve the elusive objective of measuring the pure Fe MVV Auger lineshape at the Fe/CoO interface. Similarly to what found for other FM thin films (Fe/Cu and Fe/Ag), the MVV spectrum is composed by a weakly correlated part at high kinetic energy and by a manifold of electron correlation hole-hole resonances at lower kinetic energies. The AR-APECS investigation, interpreted with the CS model, has allowed to experimentally single out different electron correlation energies U$_\textrm{eff}$ for the different pairs of final state holes. In this way, the finding of previous investigations on the dependence of U$_\textrm{eff}$ on the total spin of the final state is corroborated \cite{gotter2012, gotter2020}. The present experiment has furthermore investigated the role of electron-electron correlation in the process of exchange-coupling at Fe/CoO FM/AFM bilayer. By using a molecular calculation for Fe-O, the contribution to the AR-APECS intensity of a relatively small amount of iron atoms bonded to oxygen has been identified in a limited region of the weakly correlated spectrum and quantified to 5\% of the total AR-APECS intensity. Such an amount of Fe bonded to oxygen does not play any relevant role in the magnetic properties of the Fe/CoO bilayer. 
Finally, the comparison of the AR-APECS spectra of the two interfaces made it possible to identify, in the FM/AFM case, the increase in the exchange component of U$_\textrm{eff}$ limited to the t$_{2g}$ orbital. Conversely, the Coulomb component increased for the e$_{g}$ orbital only. Taking into account the different degree of localization of the two orbitals, similar to a Fermi liquid for the t$_{2g}$ and to a Luttinger liquid for the e$_{g}$, it has been spontaneous to associate the variation of the exchange component of U$_\textrm{eff}$ with the exchange coupling at the FM/AFM interface, due to the substrate antiferromagnetism, and of the Coulomb component with the insulating nature of the substrate.
These results are potentially of great interest to those who are about to model the technologically relevant nanosized TM/TMO bilayers.
 
\section*{Acknowledgements}
Partial financial support through:
SIMDALEE2 Sources, Interaction with Matter, Detection and Analysis of Low Energy Electrons 2 Marie Sklodowska Curie FP7-PEOPLE-2013-ITN Grant number 606988,
PRIN-2015 project NEWLI of the MIUR and ELETTRA users support is greatly acknowledged.
The authors are grateful to the ALOISA beamline team for the support received while performing the experiment at the ELETTRA synchrotron radiation source.

\begin{frontmatter}
\title{{\bf SUPPLEMENTARY MATERIAL}\\Electron correlation effects in the exchange coupling at the Fe/CoO/Ag(001) ferro-/antiferro-magnetic interface.}
\end{frontmatter}

\setcounter{section}{0}
\section{The Fe-O molecular model}
For the Fe/CoO system, a computational molecular model has been used to quantify the effect on the Fe M$_3$VV lineshape of the iron atoms bonded to the oxygen. As shown in Fig.3 of the main text, XPS measurements have allowed to quantify the formation of iron-oxygen bonds at the interface, as largely reported in literature~\cite{luches2012, colonna2009, giannotti2016, valeri2007}. The FeO molecular model has been implemented imposing a 2 \AA{}  Fe-O interatomic distance: a value that is similar to those typically found in Fe oxides \cite{alfredsson2004, hill1979, wright2002}. This calculation made use of the SURPRISES code~\cite{taioli2009, taioli2010, taioli2015, taioli2009b}  which implements a theoretical method whose main features are: (i) the double ionization is treated as a two-step process, i.e. the primary photoemission and the secondary Auger decay; (ii) the Fano's theory of the discrete-continuum interaction is used for calculating both the primary photoemission and the Auger decay; (iii) the continuum and discrete wavefunctions are calculated by means of a basis set of gaussian functions.
The chosen gaussian basis set contains the cc-pVQZ for the Fe atom and the cc-pVDZ for the O atom~\cite{calculationdetail, dunning1989}. The electronic correlations are (partially) taken into account by performing a configuration interaction (CI) calculation for the bound states and a many-channel interaction for the continuum escaping electron. The outcome of this molecular model is a remarkable number of Auger multiplet components (15 due to decay of the 3$\sigma$ core hole with $m = 0$ and 18 due to the 3$\pi$ with $m = 1$). The full set of multiplet terms contributes to a lineshape mainly regrouped in the energy interval 38 - 48 eV with an almost negligible contribution in the range 23 - 28 eV, which is an energy range not explored by the present experiment. The convolution of this rich multiplet structure with the overall broadening of 1.6 eV FWHM discussed in the main text, and an intensity set to 10\% of the overall Auger intensity, as suggested by the XPS measures, yields the dotted blue lines in Fig.~\ref{Fig4}. In this case no adjustment was applied to the energy scale of the calculation. The main contribution of such a multiplet calculation is in the region between 43 and 40 eV. 

\section{Conventional AES analysis}
The kinetic energy of an Auger electron is dictated by the binding energies of the electrons involved, the work function to extract an electron from the surface, and the effective correlation energy U$_\textrm{eff}$ ~\cite{weightman1982}. The limitation in using conventional Auger electron spectroscopy (AES) to provide precise and specific values of U$_\textrm{eff}$ is evidenced in Fig.~\ref{Fig4}a: (i) the subtraction of an underlying background becomes arbitrary when the intensity due to the secondary electrons emission is much greater than the tiny Auger peak (see inset); (ii) the Co M$_{23}$VV Auger line at about 50 eV overlaps the Fe one and the task of investigating the electronic structure of the Fe overlayer alone is hopeless; (iii) also the Fe M$_{2}$VV transition, even if with less intensity, overlaps the Fe M$_{3}$VV; (iv) finally, AES spectra, as resulting after a tentative background subtraction, are featureless.

At first glance, AES spectra could be understood as band-like, to be interpreted neglecting hole-hole correlation in the Auger final state, thus predicting the MVV Auger lineshape by the self-convolution of the Fe valence-band density of states (SCDOS). The experimental DOS for the Fe film is obtained from the photoelectron spectra of the valence band acquired with 950 eV photons: it is usually assumed that such a XPS spectrum, averaging over a larger portion of the Brillouin zone with respect to low energy UPS, better reflects the total DOS of the valence band. The experimental Fe DOS on the CoO substrate has thus been obtained after the subtraction of the substrate contribution (measured in a prior measurement) and of a Shirley background ~\cite{shirley1972}. The asymmetry of the photoline, due to the interaction of the photohole with the conduction electrons, is taken into account by a Doniach-Sunjic (D-S) profile with an asymmetry parameter $\alpha$ = 0.2~\cite{hochst1976}. The DOS retrieved in this way is subsequently self-convoluted, broadened by the experimental energy resolution, and aligned with the experimental AES peak: an alignment at the Fermi level would be more straightforward, but there is no chance to identify it for the experimental Fe M$_{3}$VV spectrum, because its onset is obscured by the overlap with Fe M$_{2}$VV transition, and even worse when also the Co M$_{23}$VV overlaps.
A similar procedure has been adopted for the Fe/CoO system, and adding to the SCDOS the contribution, concentrated in the region between 43 and 40 eV, due to the iron-oxide as calculated by the Fe-O model. One could speculate that a major difference in the AES intensity between Fe/CoO and Fe/Ag is indeed in this region, but the arbitrariness of the background subtraction does not allow to rely on the curves of the AES peak: the slope in this region is in fact strongly affected by the background subtraction.

In conclusion, AES lineshapes could be explained by the SCDOS, corrected, in the Fe/CoO case, for the contribution of the small amount of iron atoms bonded to the oxygen, as displayed by the red dashed (Fe/Ag) and the blue dashed-dotted (Fe/CoO) lines. This finding would lead to the conclusion that in these Fe thin films electron correlation is irrelevant.

\section{Angle integrated APECS spectra}
\begin{figure}[h!]
\begin{centering}
\includegraphics[width=1\textwidth]{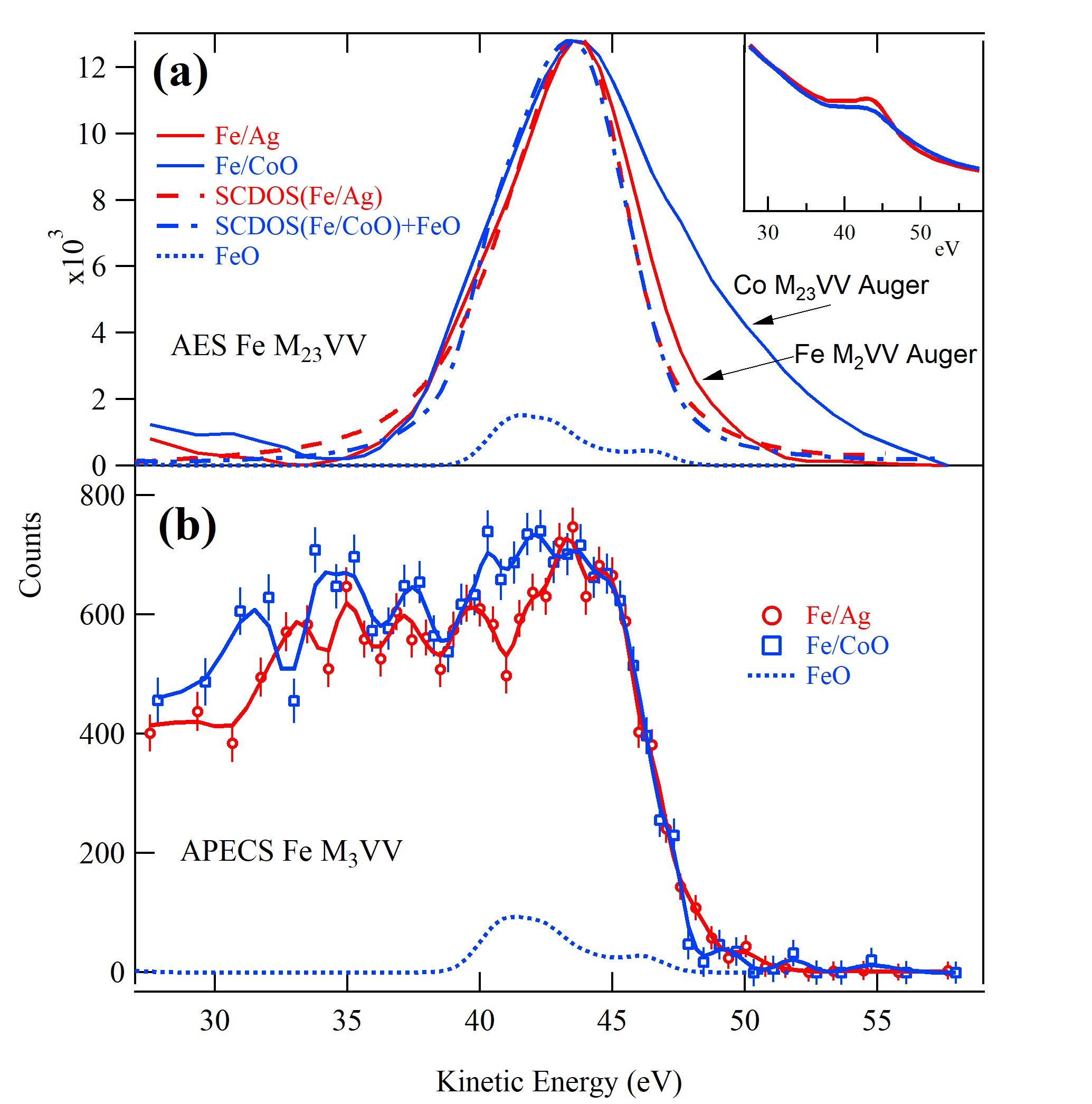}
\caption{(a): conventional AES spectra from Fe/Ag (red continuous line) and from Fe/CoO (blue continuous line) after a tentative subtraction of an integral background  from raw spectra shown in the inset plot. Fe/Ag AES is compared with the SCDOS as experimentally estimated from valence band XPS (red dashed line), while Fe/CoO AES is compared with the sum of the estimated SCDOS and the contribution due to FeO (blue dotted line), evaluated to be equal to 10 \% of the total AES intensity.  (b): Angle-integrated APECS spectra of the M$_3$M$_{4,5}$M$_{4,5}$ (3p$_{3/2} \rightarrow$ [3d; 3d]) Auger transition. The Fe/CoO experimental spectrum (square markers with error bars) is compared with the Fe/Ag experimental spectrum (circular markers with error bars). The continuous lines are guides for the eyes. The Auger intensity computed for the FeO molecule (blue dotted line) and contributing to the 5\% of the total APECS intensity, has its maximum intensity in the same 40-42 eV interval, where also the difference between the APECS spectra from the two samples is significant.}
\label{Fig4}
\end{centering}
\end{figure}

In  Fig.~\ref{Fig4}b, the two APECS spectra  for Fe/CoO and Fe/Ag, are shown. They have been collected adding up all the recorded Auger-photoelectron coincidence events, regardless of the analyzer pair; in this way, the selectivity in orbital and spin moment associated to the selection in angle of the two electrons is mitigated to the level that it becomes irrelevant~\cite{gotter2005}. Hence, APECS spectra are the pure Auger lineshapes of the two overlayers under comparison.

The improvement achieved on the Auger spectra by the time-coincident detection of Auger and photoelectrons is evident: in APECS spectra the background is completely eliminated before the Auger onset, because Co M$_{23}$VV and Fe M$_{2}$VV Auger electrons, and other prior secondary electrons cannot be detected in coincidence with the Fe 3p$_{3/2}$ photoelectrons; only an intrinsic background remains, which can be easily simulated by a Shirley background~\cite{shirley1972, tougaard1989} whose intensity is determined by the effective sampling depth of the experiment that equals the inelastic mean free path of the electron pair~\cite{satyala2014,werner2005}.
The APECS spectra are no longer featureless and span over a wider energy range, from a well defined onset, down to 30 eV of kinetic energy. It is interesting to notice that now, in the region between 40 and 43 eV, the amount by which the blue curve (Fe/CoO) is above the red one (Fe/Ag) is statistically significant and accounted for by imposing the contribution provided by the Fe-O model to 5\% of the total integrated intensity, as suggested by the coincident photoelectron analyzers setting (Fig.3 of the main text).
This confirms the already proposed existence of oxidized atoms at the interface~\cite{luches2012, colonna2009, valeri2007}. As a consequence, being the correlation in the localized FeO bonds already taken into account by the molecular calculation, the differences in the range 30 - 40 eV have to be ascribed to electron correlation in the metal iron, acting on the energy position of the sharp features as described in the main text.

\end{document}